\documentclass[sigconf, authorversion=true]{acmart}

\usepackage{balance} 
\usepackage{microtype}
\usepackage{float}
\usepackage{stfloats}

\AtBeginDocument{%
  }

\copyrightyear{2025}
\acmYear{2025}
\setcopyright{cc}
\setcctype{by}
\acmConference[CHI '25]{CHI Conference on Human Factors in Computing Systems}{April 26-May 1, 2025}{Yokohama, Japan}
\acmBooktitle{CHI Conference on Human Factors in Computing Systems (CHI '25), April 26-May 1, 2025, Yokohama,
Japan}\acmDOI{10.1145/3706598.3714218}
\acmISBN{979-8-4007-1394-1/25/04}

\makeatletter
\@ifundefined{textls}{\long\def\textls{\@ifnextchar[{\@textls}{\@textls[]}}
\long\def\@textls[#1]#2{#2}}{}
\makeatother

\begin{document}
\title[Understanding How Seams Between Virtual and Real Identities Engage VTuber Fans]{I Stan Alien Idols and Also the People Behind Them: Understanding How Seams Between Virtual and Real Identities Engage VTuber Fans – A Case Study of PLAVE}

\author{Dakyeom Ahn}
\orcid{0009-0004-9139-1605}
\authornote{Both authors contributed equally to this work.}
\affiliation{%
  \institution{Information Science and Culture Studies}
  \institution{Seoul National University}
  \country{Republic of Korea}
}
\email{adklys@snu.ac.kr}

\author{Seora Park}
\orcid{0000-0001-7281-4538}
\authornotemark[1]
\affiliation{
  \institution{Department of Informatics}
  \institution{Indiana University Bloomington}
  \country{USA}
}
\email{seorpark@iu.edu}

\author{Seolhee Lee}
\orcid{0009-0003-3273-4455}
\affiliation{%
  \institution{Department of Communication}
  \institution{Seoul National University}
  \country{Republic of Korea}
}
\email{seolheelee@snu.ac.kr}

\author{Jieun Cho}
\orcid{0009-0006-1604-8000}
\affiliation{%
  \institution{Information Science and Culture Studies}
  \institution{Seoul National University}
  \country{Republic of Korea}
}
\email{jieun99@snu.ac.kr}

\author{Hajin Lim}
\orcid{0000-0002-4746-2144}
\authornote{Corresponding author}
\affiliation{%
  \institution{Department of Communication}
  \institution{Seoul National University}
  \country{Republic of Korea}
}
\email{hajin@snu.ac.kr}

\renewcommand{\shortauthors}{Dakyeom Ahn*, Seora Park*, Seolhee Lee, Jieun Cho, Hajin Lim}

\begin{CCSXML}
<ccs2012>
 <concept>
       <concept_id>10003120.10003121.10011748</concept_id>
       <concept_desc>Human-centered computing~Empirical studies in HCI</concept_desc>
       <concept_significance>300</concept_significance>
       </concept>
 </ccs2012>
\end{CCSXML}

% \ccsdesc[300]{Human-centered computing~Empirical studies in HCI}
% \keywords{VTuber, VTubing, virtual idol, live streaming, virtual identity, seam}
\ccsdesc[300]{Human-centered computing~Empirical studies in HCI}

%% Keywords. The author(s) should pick words that accurately describe
%% the work being presented. Separate the keywords with commas.
\keywords{VTuber, VTubing, virtual idol, live streaming, virtual identity, seam}

\begin{abstract}
Virtual YouTubers (VTubers) have recently gained popularity as streamers using computer-generated avatars and real-time motion capture to create distinct virtual identities. While prior research has explored how VTubers construct virtual personas and engage audiences, little attention has been given to viewers’ reactions when virtual and real identities blur—what we refer to as ``seams.'' To address this gap, we conducted a case study on PLAVE, a popular Korean VTuber Kpop idol group, interviewing 24 of their fans. Our findings identified two main sources of seams: technical glitches and identity collapses, where VTubers act inconsistently with their virtual personas, revealing aspects of their real selves. These seams played a pivotal role in shaping diverse fan engagements, with some valuing authenticity linked to real identities, while others prioritized the coherence of virtual personas. Overall, our findings underscore the importance of seams in shaping viewer experiences.
\end{abstract}

\begin{teaserfigure}%[!ht]
  \centering
  \includegraphics[width=400px]{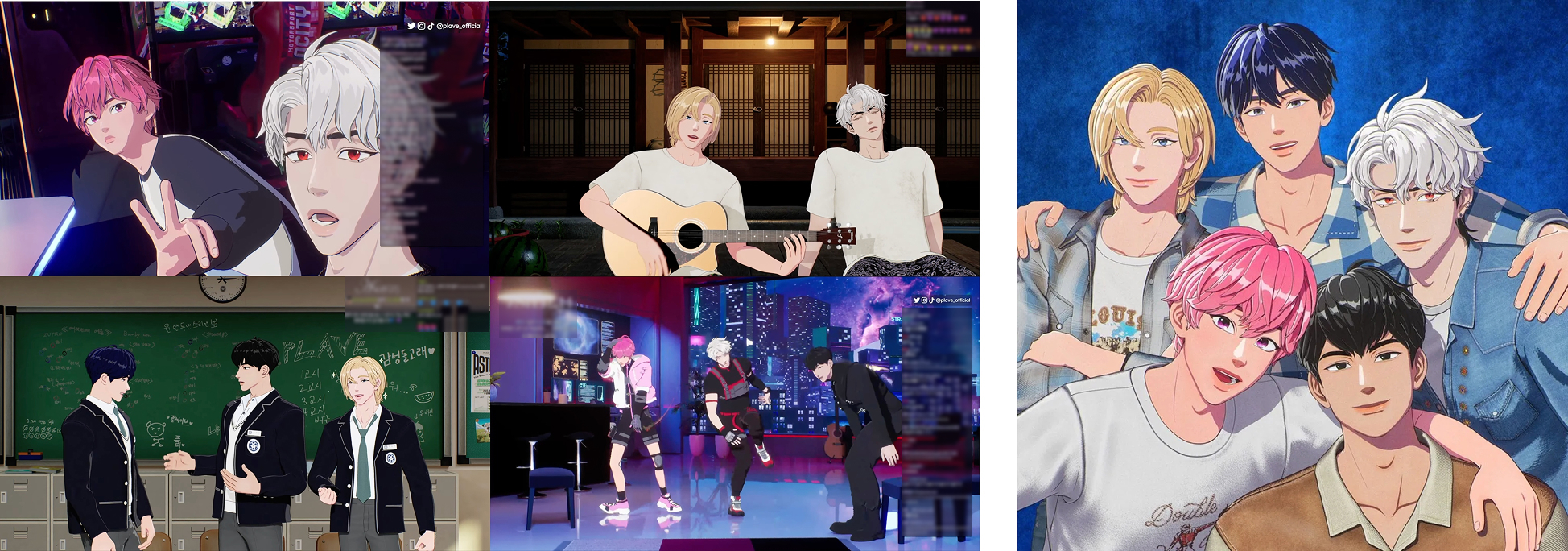}
  \captionsetup{justification=centering}
  \caption{Live streams of PLAVE (left) and the concept photo of the five members (right) ©VLAST}
  \Description{From left to right: four screenshots of PLAVE's weekly live streams on the PLAVE's Youtube channel and a group photo of their recent digital single `Pump Up the Volume.' The copyright for the images of PLAVE featured in this figure is held by VLAST.}
  \hfill
  \label{fig:plave}
\end{teaserfigure}

% \begin{teaserfigure}%[!ht]
%   \centering
%   \includegraphics[width=425px]{figures/teaser.png}
%   \captionsetup{justification=centering}
%   \caption{Overview of our findings: A fan’s experience of engaging with a VTuber, progressing to recognizing their real identity, and ultimately choosing to embrace both personas or focus solely on the virtual.}
%   \Description{This figure illustrates the fan engagement process with VTubers in a clockwise progression. Fans first become interested in a VTuber as a digital persona. As they engage further, they notice subtle cues hinting at the real person behind the avatar. This leads to the discovery of the real identity, where they recognize both similarities and differences between the virtual and real personas. In response, fans either embrace both identities as interconnected or detach from the real identity, choosing to focus solely on the virtual persona}
%   \label{fig:fig1}
% \end{teaserfigure}

\maketitle

\section{Introduction}
% \begin{figure*}[!hb]
%   \centering
%   \includegraphics[width=480px]{figures/fig1.jpg}
%   \captionsetup{justification=centering}
%   \caption{PLAVE’s live streams (left) and concept photo (right) of five members, \\Noah, Bambi, Yejun, Hamin, and Eunho (left to right) ©VLAST}
%   \Description{From left to right: four screenshots of PLAVE's weekly live streams on the PLAVE's Youtube channel and a group photo of their recent digital single `Pump Up the Volume'. The copyright for the images of PLAVE featured in this figure is held by VLAST.}
%   \label{fig:fig2}
% \end{figure*}

Virtual live streamers, also known as Virtual YouTubers (VTubers), are a rapidly emerging group of human content creators who interact with audiences through computer-generated 2D and 3D avatars on streaming platforms like YouTube, Twitch, and Bilibili ~\cite{wan2023investigating}. They often utilize specifically designed animation-style avatars to perform a wide range of live-streaming activities, including singing, gaming, and chatting with viewers. Over time, the VTubing domain has extended to encompass various entertaining roles, such as virtual idols, models, and DJs ~\cite{ferreira2022vtuber}.
Since its emergence in Japan in 2017, the VTuber market has grown rapidly, hosting over ten thousand active creators worldwide ~\cite{liudmila2020designing}. By 2032, the VTuber-related economy is projected to reach an estimated market value of \$498 billion~\cite{business2024}.

VTubers leverage various technologies such as real-time face and motion tracking, digital art, 3D modeling, and voice modulation to craft virtual personas that may differ in gender, appearance, and personality, transcending their real-life identities ~\cite{wan2023investigating, ferreira2022vtuber, lu2021more,chen2024conan}.
They even enact unconventional characters, including aliens, animals, and anime figures ~\cite{lu2018you, hamilton2014streaming, li2021drives}. 
This `virtual identity,' encompassing the digital avatar and its fictional narrative, is usually distinct from the `real identity' of the person operating it ~\cite{wan2023investigating}. VTubers perform these virtual identities during live streams as if they were authentic entities by mapping their real movements and voices onto their virtual characters. 

This unique dual identity structure embodied in VTubers significantly influences how fans engage with and consume VTuber content. Recent studies have revealed that most VTuber fans develop strong emotional connections with the virtual representations of the avatar, often without knowing the real identities behind them ~\cite{freeman2024myaudience, lu2021more, brett2022we,liudmila2020designing}. 
This focus on virtual identity typically leads to a shared norm within the VTuber viewership and fandom community that prohibits revealing the real identities of VTubers. This community norm serves not only to protect the creators' privacy but also to help maintain viewers' immersion~\cite{lu2021more, chen2024conan}.

\begin{figure*}[!hb]
  \centering
  \includegraphics[width=416px]{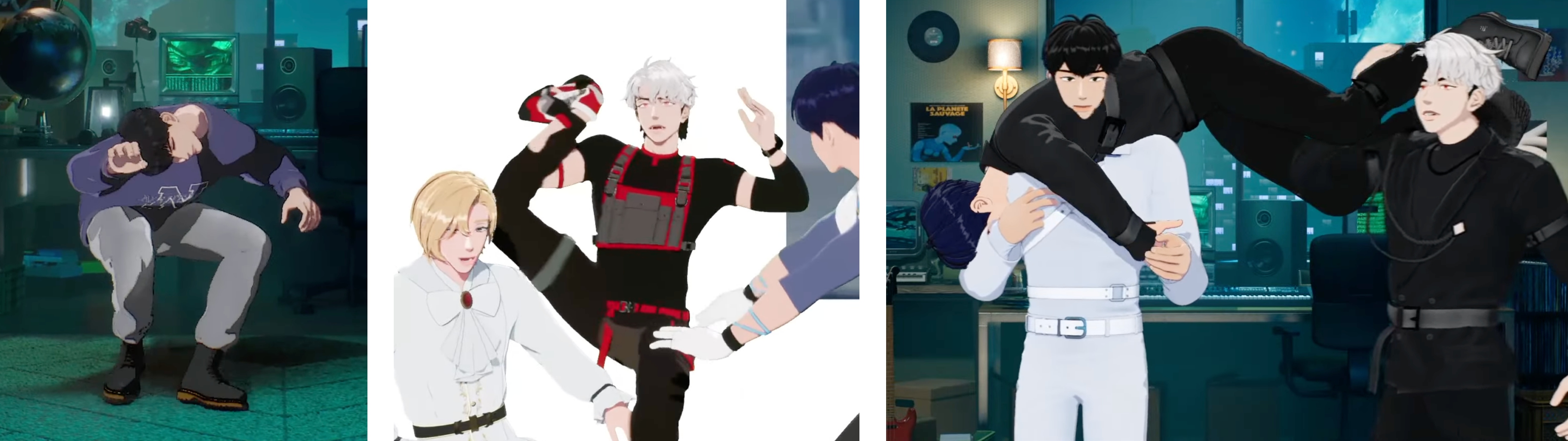}\captionsetup{justification=centering}
  \caption{Technical glitches during PLAVE's live streaming. Members' necks and legs were twisted (left \& center). \newline They even floated in the air, penetrating another member's head (right). ©VLAST}
  \Description{The three screenshot images illustrate the moments where technical glitches during PLAVE's live streaming. In the left image, Hamin's neck appears twisted unnaturally. The middle screenshot captures Eunho with a distorted leg and his arm seemingly penetrating through it. The right image presents Yejun floating in the air, with parts of his body intersecting Hamin's head. The copyright for the images of PLAVE featured in this figure is held by VLAST.}
  \label{fig:glitches}
\end{figure*}

However, VTubers often face challenges in consistently maintaining their virtual identity due to the nature of live streaming. The high level of interactivity and real-time engagement inherent to live streams often results in moments where the virtual identity becomes disrupted, exposing elements of the performer’s real identity ~\cite{haimson2017makes, hamilton2014streaming}. These risks are further heightened by the inability to edit or reverse live-streamed content ~\cite{lampinen2011we, li2018tell,bodenheimer1997process}. As a result, there have been several incidents of accidental identity exposure of VTubers, such as Oshiro Mashiro (@mashiro0529) inadvertently reflecting their face on a Nintendo Switch screen during a stream ~\cite{tamaki2022deletion}, and Filian (@filianIsLost) accidentally revealing their face while googling themselves ~\cite{silkmagazine2024filian}. 

These incidents showed that the clear separation between the virtual and the real becomes difficult within the VTubing ecosystem, where the boundary between human performers and their avatar characters often becomes blurred. This blending creates a liminal space that invites deeper exploration into how viewers interact with VTubers' incongruent identities. To better understand the nuanced interplay between the real and the virtual in VTuber-fandom relations, we employ the concept of `seam'—a point where the constructed virtual identity of VTubers becomes inconsistent or breaks down, revealing aspects of the real human performers behind the virtual avatars.  
Similarly, previous discourses on virtual environments have framed seams as moments of complexity, ambiguity, or inconsistency that disrupt the `seamless' integration of the real and the virtual ~\cite{inman2019beautiful, coleman2011hello, lin2024actualities, Garcia2017seamless}. Building on this framework, this study aims to investigate how these incongruous moments, referred to as seams, emerge in the VTubing context and shape viewer perceptions and engagement.

To explore how VTuber fans recognize and interpret seams, we chose PLAVE as a case study. PLAVE is a popular five-member K-pop VTuber idol group that debuted in South Korea in 2023. PLAVE utilizes real-time motion capture technology and Unreal Engine to implement 2D avatars in 3D virtual environments ~\cite{unrealenginekr_2023}. PLAVE primarily engages with audiences through live streams on YouTube, featuring dance performances, singing, gaming, and songwriting sessions, alongside casual live chats with viewers (see Figure \ref{fig:plave}).

We chose PLAVE as our case because their content exemplifies the dynamic interplay between virtual and real identities, particularly through the frequent and visible seams in their live streams. A notable aspect of this case is that PLAVE has garnered attention through compilation videos highlighting technical glitches~\cite{lee2024plave}. These videos feature moments when motion capture technology malfunctions during live streams, resulting in uncanny and unrealistic distortions in their avatar bodies (see Figure \ref{fig:glitches}). While these glitches reveal the imperfection of the crafted virtual persona, the majority of viewers perceive these moments as playful, amusing, and entertaining ~\cite{park2024korea,hankyoreh2024}. PLAVE's sustained fan engagement, even in the face of these seamful exposures at times, makes them an ideal case for examining how seams shape VTuber viewership and fandom dynamics.

Building on these observations, we sought to further explore how seams influence viewers' interactions with VTubers by interviewing 24 PLAVE fans. To guide our investigation, we posed the following research questions (RQs):

\begin{itemize}
\item\textbf{RQ1.} What situations lead viewers to recognize seams in VTubers?
\item\textbf{RQ2.} How do VTuber viewers perceive and interpret these seams? 
\item\textbf{RQ3.} How do viewers engage with these seams in their interactions with VTubers?
\end{itemize}

Our findings revealed that as participants became interested in and engaged with PLAVE’s content, they began to recognize seams through moments when the virtual identity as alien idols was disrupted by technical glitches or subtle cues revealing the performers’ real identities. These seamful moments sparked curiosity about the individuals behind the avatars. Upon acknowledging the presence of real people behind these virtual identities, participants adopted two distinct approaches to incorporating them into their engagement. More than half of the participants integrated both the virtual and real aspects to form a cohesive understanding of PLAVE, while others deliberately distanced themselves from the real identities, focusing solely on the virtual personas.

The contributions of this work are as follows: First, we comprehensively mapped out the process through which viewers recognize, interpret, and integrate seams into their interactions with VTubers. Second, we contribute to broader discussions on identity and authenticity, offering new insights into the evolving landscape of virtual entertainment. Finally, we discuss design implications for platforms to better accommodate diverse viewer expectations and experiences regarding seams.

\section {Research Background}
% This section explores how VTubers operate their dual identities—balancing curated virtual personas with glimpses of their real selves—and the impact this interplay has on audience engagement. We introduce the concept of `seams,' moments where these identities blur, disrupting immersion but offering opportunities for a new interpretation. By examining existing research and the case of PLAVE, a Korean VTuber K-pop idol group, we highlight how seams set the stage for understanding the broader dynamics of hybrid virtual-real interactions.

In this section, we first review the existing literature on VTubers' dual identities through the lens of Goffman’s dramaturgical approach ~\cite{goffman2023presentation}. Next, we examine the concept of `seams' in prior HCI research, establishing the foundation for our study. Finally, we introduce our case study, PLAVE, and explain why it provides a compelling context for this research.

\subsection{VTubers' Dual Identity and Viewer Engagement} 
Virtual YouTubers (VTubers) are live streamers who present virtual identities in their content using 2D or 3D digital avatars ~\cite{Shenshen2021}. These streamers interact with their audience through real-time live streams, engaging in activities like singing, dancing, gaming, cooking, and having casual conversations. All of these activities are presented through virtual personas that often differ significantly from their real selves in terms of gender, characteristics, appearance, and voice ~\cite{bredikhina2022becoming,chen2024conan}. VTubers thus operate with a dual identity structure: a `virtual identity' that encompasses the digital avatar along with its associated roles, personalities, and fictional backstories, and a `real identity' that belongs to the human performers who control the avatars' behaviors and performances from behind the scenes ~\cite{wan2023investigating}.

In navigating this duality, Erving Goffman’s dramaturgical approach offers a useful framework for analyzing the interactions between VTubers and their viewers, particularly his concepts of `front stage' and `backstage' ~\cite{goffman2023presentation}. In this context, the `front stage' represents the public persona that VTubers curate to meet audience expectations or conform to platform norms. The `backstage' is where these performers step away from their virtual roles, revealing more candid aspects of their personalities or behaviors.

To establish their virtual identities at the `front stage,' VTubers employ a range of animating and modeling technologies, notably motion capture technology that tracks performers' body movements and facial expressions and converts them into animations of the virtual character ~\cite {guarriello2019never, cakir2024become}. Their voices are also often modified to match specific vocal features of the virtual identity
~\cite{chen2024conan}. Previous work has found that these physical attributes, notably anime-like appearances, played a vital role in attracting viewers ~\cite{li2023does,brett2022we,liudmila2020designing}. 

At the same time, some VTubers gradually disclose aspects of their `backstage' selves, shaping their virtual personas to feel more original and authentic. To strengthen their connection with viewers, some VTubers strategically share personal anecdotes or humorously reference their real selves during live streams ~\cite{wan2023investigating, wijaya2023language}. Furthermore, others deliberately weave elements of their real identities into their virtual personas, creating more layered, multidimensional characters ~\cite{tang2022dare, pellicone2017game, wan2023investigating}.

% This initial attraction is deepened through interactive activities during live streams, such as singing along at virtual concerts and creating memes with the fan community ~\cite{lu2018you, lee2019understanding,lee2023ju}. This bilateral engagement subtly implies VTubers' `backstage' personalities, helping them develop their virtual personas to be more original and foster closer relationships with their viewers. Therefore, some VTubers strategically share personal information or make fun of their real selves during VTubing ~\cite{wan2023investigating,wijaya2023language}. Furthermore, some VTubers are known to infuse aspects of their real selves into their virtual identities, thereby enriching their presentations and creating more complex personas ~\cite{tang2022dare, pellicone2017game, wan2023investigating}.

However, this blending of front stage and backstage can create tension. Since VTuber viewers often value a consistent portrayal of virtual identities and tend to distance themselves from the performers’ real identities ~\cite{lu2021more}, any exposure to the `backstage' can disrupt viewers' sense of immersion. As a result, VTuber fan communities often establish strict norms that discourage or outright prohibit discussions about the real individuals behind the avatars ~\cite{wu2022concerned,wijaya2023language}. Despite these efforts, maintaining a seamless virtual identity during live streams is challenging due to the highly interactive and synchronous nature of the medium, which can lead to unintended disclosures of the real identity through mishaps or deviations from the constructed virtual persona ~\cite{haimson2017makes, hamilton2014streaming, wan2023investigating}.

As such, the intentional or accidental blending of virtual and real identities in VTubing creates a dynamic interplay that influences audience engagement. However, how viewers recognize and react to this interplay, particularly within the VTubing context, remains largely unexplored. Therefore, gaining a deeper understanding of how viewers interpret and navigate these seams is crucial for revealing the broader implications of virtual-real identity interplay, including its impact on audience engagement.

% The intentional or unintentional blurring of virtual and real identities in VTubing creates a unique interplay that shapes how viewers engage with VTubers. However, little is known about how audiences perceive and respond to this blending, especially within the VTubing context, leaving a gap in understanding its impact on viewer experience and engagement. These moments, referred to as `seams,' occur when VTubers' virtual identities collapse or blend with their real selves. Understanding how viewers navigate these seams is crucial for uncovering the dynamics of virtual-real identity interplay.

\subsection{Seams in VTubing Experiences}
The concept of `seams' has been discussed in early discourses on ubiquitous computing as points where the boundaries between systems, interfaces, or realities become visible and enter users' awareness~\cite{broll2005seamful, rubambiza2022seamless, inman2019beautiful}. In contexts where technology is integrated into everyday life, the visibility of seams has often been viewed as disrupting smooth interactions, hindering the technology from being fully embodied within users' contexts ~\cite{weiser1991computer}. 

Specifically, in studies of simulated environments such as virtual and mixed reality and the metaverse, seams are often characterized as mismatches, gaps, or inconsistencies between the virtual and physical worlds~\cite{broll2005seamful, rubambiza2022seamless}. They focused on how seams interrupt immersive experiences, causing breakdowns in users' engagement~\cite{grimshaw2014oxford, damer2014virtuality}. As such, seams have been framed as obstacles to user experience, prompting designers to prioritize eliminating or concealing them to create seamless systems~\cite{Hengesbach2022undoing, inman2019beautiful}. Consequently, much of the literature emphasizes the importance of making seams invisible to achieve smooth integration between the user’s reality and simulated environments~\cite{coleman2011hello, lin2024actualities, Garcia2017seamless, freeman2024myaudience}.

While seamlessness has traditionally been valued in system design~\cite{inman2019beautiful}, an alternative approach, known as `seamful design,' has gained traction in recent years~\cite{chalmers2003seamful, broll2005seamful, chalmers2004social}. This perspective emphasizes ``seamful moments''—instances of breaks, gaps, or misalignments in user experiences—as opportunities to uncover hidden struggles and negotiations that are often overlooked by designers or researchers~\cite{bell2006interweaving, lyytinen2002ubiquitous, inman2019beautiful}. For example, Rubambiza and colleagues \cite{rubambiza2022seamless} showed how seams in digital agriculture infrastructure exposed gaps in user expectations, creating both challenges and opportunities for improvement. Similarly, Erickson and Jarrahi \cite{erickson2016infrastructuring} demonstrated how breakdowns in ICT infrastructure prompted users to develop strategies that enhanced their understanding and agency. Also, Dao and colleagues~\cite{dao2021bad} leveraged moments of user failure in VR technology as strategic points to enhance user engagement.

Furthermore, seams can serve as a reflective tool for users, helping them interpret embedded glitches and navigate their relationships with technology~\cite{vertesi2014seamful, sengers2006staying}. By recognizing seams, users can reason about causality, develop a deeper understanding of the system, and adjust their behavior accordingly~\cite{kratz2009unravelling, erickson2016infrastructuring}. Notable studies in VR contexts have highlighted these dynamics. For example, O’Hagan and colleagues \cite{o2021safety} found that the physical presence of bystanders enhanced VR users’ immersion by fostering awareness of their surroundings. Similarly, researchers have introduced intentional interruptions—such as verbal communication, notifications, and visual cues—to support users in transitioning between physical and virtual realities\cite{ghosh2018notifivr, o2023re, kudo2021towards}.
These design approaches align with the principles of seamful design, which ``\textit{involves deliberately revealing seams to users and taking advantage of features usually considered as negative or problematic}'' ~\cite[p.1]{chalmers2003seamfulubicomp}. 
% In this way, seams facilitate transitions between physical and simulated environments, turning what might initially appear as disruptions into meaningful interactions.

Building upon the rich discourse on seams in HCI, we extend the concept of seams to the domain of VTubing, a context where hybrid interactions between physical and virtual realities are central. In line with earlier discussions of seams as points of mismatch or disruption, we define seams in VTubing as moments when the seamless representation of a virtual identity break down. These moments can emerge from technical glitches, inconsistencies, or implicit revelations of the real identities behind the avatars, akin to the gaps and breakdowns discussed in prior research on VR and mixed reality environments~\cite{broll2005seamful, rubambiza2022seamless, dao2021bad}.

In VTubing, such seamful moments blur the boundaries between reality and virtuality, interrupting the immersive experience while also creating opportunities for reinterpretation. This perspective aligns with the seamful design approach, which views visible seams not as mere obstacles but as openings for users to engage with underlying complexities~\cite{chalmers2003seamfulubicomp, vertesi2014seamful}. By exposing the duality of VTubers’ virtual and real identities, seams can shape the viewer’s experience in unique and meaningful ways. While previous research has primarily focused on the challenges VTubers face in managing their dual identities~\cite{wan2023investigating, lu2021more}, our study shifts the focus to how these seams function as critical elements that shape viewer engagement.

\subsection{Case Study: PLAVE}
To explore how seams shape viewer engagement, we investigate the case of PLAVE, a popular Korean VTuber idol group. PLAVE is a virtual idol group from South Korea comprising five members: Yejun, Noah, Bamby, Eunho, and Hamin (see Figure \ref{fig:plave}). Utilizing advanced Unreal Engine and motion capture technology, they animate webtoon-style 2D avatars in simulated environments. Specifically, the group’s conceptual narrative positions them as forgotten webtoon characters from a virtual universe called \textit{Caelum}, interacting with fans on Earth through an intermediary realm known as \textit{Asterum}. This fantastical setup merges digital imagery with surreal narratives, encouraging fans to immerse themselves in the unique characters and narratives of this group ~\cite{park2024korea}. A distinctive aspect of PLAVE’s appeal is their use of 2D avatars via 3D modeling, which contrasts with the hyper-realistic 3D visuals typical in K-pop (e.g., nævis from SM Entertainment~\cite{kim2024naevis}).

PLAVE regularly engages with fans through weekly live streams on their YouTube channel\footnote{\url{https://www.youtube.com/@plave_official}}, where they share diverse content, including singing, dancing, songwriting, gameplay, and personal updates. Their live streams have attracted an average of over 15 thousand viewers per episode, with peak audiences reaching over 30 thousand ~\cite{park2023story}. In addition to their digital presence, PLAVE has achieved considerable success in the Korean music industry, further solidifying their position as a virtual idol group comparable to conventional K-pop idols. Their fanbase is predominantly female~\cite{concert2024}, consistent with the global K-pop fandom demographic, where more than 90\% of fans are women~\cite{HanteoGlobal2021}. They actively perform pop culture content, including K-pop songs, dance performances, music videos, and albums, bridging the gap between traditional and virtual idol groups. Their songs have surpassed over 10 million views on YouTube~\cite{plave_6th_summer, plave_way_4_luv} and reached first place on Korea’s most prominent music chart~\cite{kim_2024}. Notably, their recent album, `ASTERUM: 134-1,’ sold more than 500 thousand copies in its first week, showcasing their commercial appeal and the growing acceptance of VTubers in mainstream K-pop culture~\cite{hanteo2024asterum}. 

PLAVE’s unique position in the VTubing space offers various empirical examples of the interplay of virtual and real identities within fandom engagement. A notable case is fandom's reaction to technical glitches during live streams (see Figure \ref{fig:glitches}), where members' bodies appear as twisted, bent, or broken due to malfunctions of motion capture technology. Despite its unrealistic, even uncanny appearance, the glitch compilation videos\footnote{\url{https://youtu.be/srlbue3glVs?si=GL52-Gl0RLTEiyiA}} have inadvertently garnered significant popularity on YouTube, amassing 1.2 million views ~\cite{plaving2023}. 

\begin{figure}
  % \centering
  \includegraphics[width=240px]{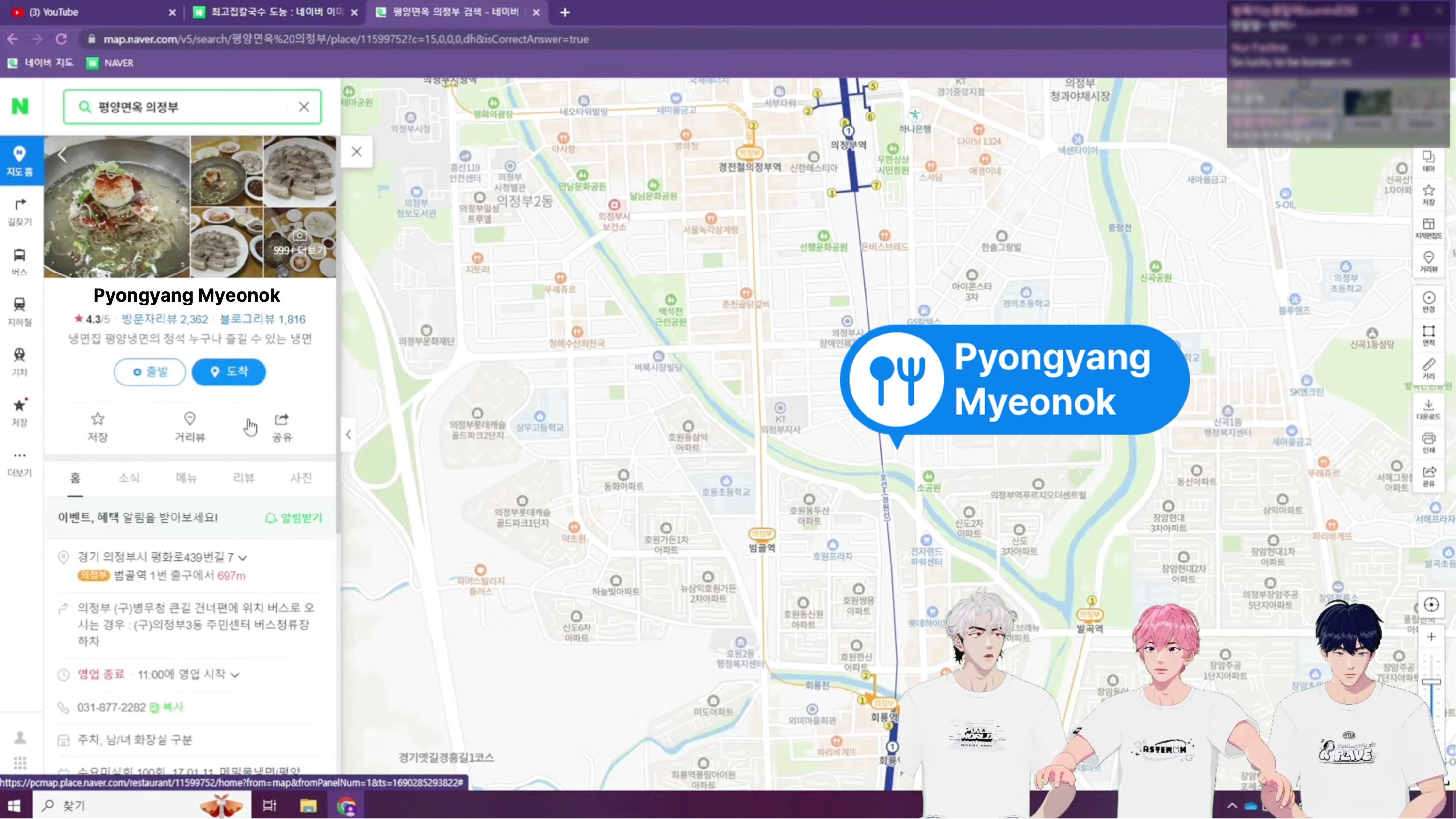}
  \captionsetup{justification=centering}
  \caption{PLAVE recommending their favorite restaurant in Seoul during live streaming ©VLAST}
  \Description{Screenshot demonstrates that PLAVE members favor a specific Korean restaurant, 'Pyongyang Myeonok.' The image shows PLAVE members Eunho, Bamby, and Yejun searching for Pyongyang Myeonok on an online map during a live streaming session, recommending it as one of their favorite restaurants in Korea. }
  \label{fig:realcue}
\end{figure}

Although PLAVE’s agency prohibits disseminating performers’ real identities to protect them from potential harm, elements of their real lives occasionally emerge through casual interactions on platforms such as YouTube live streams and celebrity-fan communication platforms, such as ` Bubble’\cite{dearu_bubble}. While maintaining their virtual alien personas, PLAVE engages in friendly and candid communication with fans. For example, members have explicitly recommended popular Korean restaurants in Seoul—despite their fictional roles as aliens living in \textit{Caelum} (see Figure \ref{fig:realcue}). Additionally, diverse public sources and social media reveal the human performers' appearances and pre-debut histories, including their struggles to become K-pop idols and their eventual success through virtual debuts \cite{um2024plave}.

These glitches and real-world references illustrate the unique nature of VTubing experiences within PLAVE fandoms, which exist at the intersection of everyday life and fantastical narratives. By blending elements of the real and the virtual, PLAVE provides a compelling case for examining how these seamful moments influence and shape viewers' engagement, fostering dynamic and multifaceted interactions.

\section{Research Method}
To explore how seams shape viewers' engagement and immersion in VTubers, we conducted semi-structured interviews with 24 fans who actively engage with PLAVE’s content. Given the complex and relatively underexplored nature of the research topic, semi-structured interviews were chosen to uncover nuanced experiences and perspectives~\cite{kvale2009interviews}. 

% This approach allowed us to explore participants’ unique insights while maintaining the flexibility to follow up on specific responses. It provided rich contextual data that went beyond what surveys or structured interviews could achieve ~\cite{kvale2009interviews}. 
% By analyzing the interview data using thematic analysis ~\cite{patton2014qualitative}, our findings reveal how participants recognized, interpreted, and engaged with these seamful moments.

\begin{table*}
  \centering
  \caption{Participant Demographics and Fan Activity Information}
  \label{tab:participant_info}
  \begin{tabular}{l c c l || l c c l}
    \toprule
    \textbf{ID} & \textbf{Gender} & \textbf{Age} & \textbf{Fan Activity Period} & 
    \textbf{ID} & \textbf{Gender} & \textbf{Age} & \textbf{Fan Activity Period} \\
    \midrule
    P1  & F & 27 & 1-3mo  & P13 & F & 29 & 7mo-1yr \\
    P2  & F & 22 & 1yr+   & P14 & F & 27 & 7mo-1yr \\
    P3  & F & 23 & 4-6mo  & P15 & F & 33 & 7mo-1yr \\
    P4  & F & 26 & 7mo-1yr & P16 & F & 23 & 1-3mo  \\
    P5  & F & 27 & 1yr+   & P17 & F & 20 & 1yr+    \\
    P6  & F & 25 & 7mo-1yr & P18 & F & 20 & 1yr+    \\
    P7  & F & 23 & 1-3mo  & P19 & F & 27 & 7mo-1yr \\
    P8  & F & 23 & 1-3mo  & P20 & F & 21 & 4-6mo   \\
    P9  & F & 25 & 7mo-1yr & P21 & F & 30 & 1yr+    \\
    P10 & F & 22 & 1yr+   & P22 & F & 31 & 1yr+    \\
    P11 & F & 23 & 1yr+   & P23 & F & 30 & 7mo-1yr \\
    P12 & F & 19 & 1yr+   & P24 & F & 26 & 7mo-1yr \\
    \bottomrule
  \end{tabular}
  \Description{This table displays the participant demographic and fan activity information of 24 participants in a two-column layout. Each column includes the participant's ID, gender, age, and fan activity.}
\end{table*}

\subsection{Recruitment}
To recruit PLAVE fans from diverse backgrounds and varying levels of fan engagement, we utilized multiple social media channels to reach PLAVE’s online fandom community. Specifically, we posted recruitment notices on online communities for university members (`Everytime'), as well as on platforms like `X,' `Instagram,' and group chats/message boards. The recruitment posts outlined our research goals, purpose, compensation, and eligibility criteria (adults aged 19 and older who identified themselves as PLAVE’s fans, watched more than an hour of PLAVE’s video content per week, and had participated in at least one of PLAVE’s live streams), and privacy measures. In order to participate in the study, potential participants were required to fill out a sign-up form with their pseudonyms, age, gender identity, and contact information. Participants were also asked several questions to confirm their status as PLAVE fans, including, ``When did you become a fan of PLAVE?'', ``How many hours per week do you watch PLAVE-related content on average?'', and ``How many times have you participated in PLAVE’s live streaming?'' The responses to these questions ensured that participants met the participation eligibility criteria.

\subsection{Participants}
In this study, we recruited 24 participants with an age range of 19 to 33 (M = 25, SD = 3.67). Regarding their fan engagement duration, 75\% of our participants (n=18) had engaged with PLAVE for more than six months, with half of them (n=9) following the group for over a year. The remaining 25\% (n=6) had become fans within the past 6 months. All participants in this study identified themselves as women and were residing in South Korea at the time of the study. While we aimed to recruit a more diverse sample, the demographics of our participants naturally reflected the composition of PLAVE’s predominantly female fandom. For instance, ticketing statistics from PLAVE’s most recent offline concert revealed that 98.9\% of attendees identified as women, with the majority in their 20s (56.7\%) and 30s (33.4\%)\cite{concert2024}. This pattern aligns with global demographic trends in K-pop fandoms, where 92\% of fans are women \cite{HanteoGlobal2021}.

\subsection{Study Procedure}
Before the interviews, participants signed a consent form detailing the study’s goals, procedures, privacy measures, and their rights. This document was provided in Korean, the native language of both participants and researchers. Interviews were conducted online via Zoom, also in Korean, with participants given the option to keep their cameras on or off to prioritize their comfort and privacy.

Following that, the interviews began with a brief introduction to the study and lighter questions designed to create a relaxed atmosphere before transitioning into more in-depth discussions. Participants were first asked, ``Could you describe how you first discovered PLAVE and became a fan?'' This question aimed to explore the initial motivations that led them to engage with PLAVE. They were also prompted to reflect on their prior knowledge of or interactions with other VTubers, virtual idols, and similar virtual content to compare these experiences to their engagement with PLAVE. We also asked how participants typically engaged with PLAVE’s content. 

Another key focus of the interviews was participants’ awareness of the real identities behind PLAVE's virtual avatars. For example, participants were asked, ``How aware are you of the real individuals behind PLAVE, and does this knowledge affect your fan experience?'' This line of questions helped explore how fans navigated the interplay between virtual and real identities, and whether such awareness enriched or detracted from their engagement. Throughout the discussions, they were encouraged to provide specific examples to ensure their responses extended beyond general opinions. For instance, they were invited to elaborate on memorable moments from Bubble messages, live streaming episodes, or videos if they felt comfortable sharing these details.

As the interviews concluded, participants were given an opportunity to share any final thoughts or comments they might not have expressed earlier. Each interview lasted approximately 60 minutes, and participants were compensated ₩20,000 (approximately \$14 USD). The study protocol received approval from the Institutional Review Board (IRB) of Seoul National University, where the study was hosted.

\subsection{Data Analysis}
We conducted a thematic analysis ~\cite{patton2014qualitative} of semi-structured interview data to investigate the impact of seams on the overall VTuber viewing experiences.  All interviews were audio-recorded, and any images or videos shared by participants during the sessions were digitally captured and integrated into each participant’s transcript. After removing identifiable information, four authors conducted the thematic analysis to identify recurrent codes, concepts, and high-level themes. Four authors independently reviewed each transcript multiple times and applied open coding to generate initial codes. Common themes identified by the researchers were then synthesized into high-level themes. To ensure analytic reliability, all authors collaborated to review and refine thematic clusters through multiple rounds of discussion. 

This entire analytical process was conducted in Korean to preserve the authenticity and nuances of the data. All authors, proficient in both Korean and English, translated the participant's illustrative quotes into English during the writing phase.

\section{Findings}
In this section, we first outline how participants' prior experiences with VTubing shaped their initial understanding of the virtual identities of PLAVE. Next, we detail the process by which participants recognized seams. Finally, we examine how participants engaged with these seams.

\subsection{Making Sense of the Virtual Identities of PLAVE}

Participants’ prior exposure to VTubing and virtual media played a key role in shaping their initial perceptions of PLAVE’s virtual identity. Among the 24 participants, the majority (n=17) were familiar with the concept of VTubing. Among them, five participants (P10, P12, P17–P18, P21) had some experience engaging with other VTuber content. These participants who had engaged  mentioned occasionally watching live streams, stage performances, or fan-made videos featuring popular VTubers, such as Japanese VTubers from Nijisanji (\href{https://www.youtube.com/@nijisanji}{@nijisanji}) and the Korean virtual idol group-- Isegye Idol (\href{https://www.youtube.com/@waktaverse}{@waktaverse}). Drawing on these experiences, they contextualized PLAVE as one of the VTuber K-pop groups but noted clear differences in their presentation and performance style. 

In particular, they perceived PLAVE’s use of advanced full-body avatar animation technology was particularly stood out. Unlike most VTubers, who typically rely on fixed cameras that primarily track facial movements, PLAVE’s technology allowed for seamless and fluid full-body motion, including natural hand and foot movements. This created a more human-like and dynamic performance that set PLAVE apart. As P18 remarked, \textit{``Most VTubers use a fixed camera to track their face, which makes their hand and foot movements look awkward. But PLAVE moves so naturally, even their hands and feet, that they feel surprisingly human-like.''} 

Others who were simply familiar with VTubing but had not actively engaged with it approached PLAVE with different expectations, shaped by their familiarity with virtual singers such as Vocaloids, K/DA, or Heartsteel from the online game, \textit{League of Legends}. These characters involve human input during their creation, such as voice acting and motion capture, but their performances are typically pre-recorded rather than live. Participants familiar with these examples initially assumed that PLAVE operated similarly. Some even perceived PLAVE as entirely AI-driven virtual performers until further exploration revealed the involvement of real human actors in their motion capture and performances.

In contrast, seven participants had no prior exposure to VTubers or virtual idols before discovering PLAVE, leaving them unsure about how to categorize PLAVE. These participants described initial confusion upon encountering PLAVE for the first time, unsure whether they were anime characters, AI-generated figures, or something entirely new. For example, P11 shared: \textit{``When I first saw their stage, I couldn’t figure out if they were human performers or anime characters at all.''} This lack of familiarity required these participants to spend more time understanding PLAVE’s hybrid nature as a virtual group operated by human performers.

\subsection{Recognizing Seams} 

Despite their varying initial perceptions, participants gradually developed a deeper understanding of PLAVE's unique virtual identity. Over time, they transitioned from perceiving PLAVE as purely a virtual construct to recognizing the human performers behind the avatars. A pivotal moment that solidified many participants’ interest in PLAVE’s dual identities came with the rise of glitch compilation videos. 

These videos compiled moments when the real-time motion capture tracking failed during live streaming. In these moments, members' bodies appeared to be twisted, to penetrate their thigh or another member’s neck, or to be suddenly floated in the air (see Figure \ref{fig:glitches}). These glitches flustered the members, resulting in spontaneous reactions that went viral due to their humorous nature. P10 recalled that the glitch was \textit{``extremely funny because it often resulted in absurd situations, like limbs appearing in the wrong places, necks twisting oddly, or characters levitating. These unexpected glitches were just comical.''} The humor inherent in these glitches contributed significantly to the virality of these compilation videos.

As the glitch compilation videos gained popularity, many participants reported that these technical failures ironically sparked their interest in PLAVE. Those who had previously been indifferent to PLAVE or only casually listened to their music found themselves intrigued by the glitches. As P23 mentioned, \textit{``I wasn't interested at first, but as their glitch videos trended on YouTube, I naturally became a fan.''} The playful and imperfect nature of these moments paradoxically humanized the virtual group and helped participants see PLAVE as more than just virtual characters; it piqued their curiosity about both the group and the human performers behind the avatars. P6 explained, \textit{``After watching the glitch video, I started looking into each member closely and even began listening to their music.''} Similarly, P1 noted that these videos transformed many casual viewers into engaged fans: ``\textit{When asked about what drew people to PLAVE, I'd say 90\% of fans will mention those glitch videos.}'' 

As interest grew, participants expanded their engagement with PLAVE beyond music and performances, beginning to watch their regular YouTube live streams. These streams provided subtle yet interesting glimpses into the performers’ real selves. P8 said, \textit{``At first, I thought their concept was extraordinary. But as I watched more live streams, I realized how passionate they are about writing and composing their own songs. They’re energetic and nice people, which made me curious about them more.''} Similarly, pre-debut stories shared during streams grounded the group's fantastical narratives in sincerity and dedication, even though these disclosures occasionally diverged from their virtual concept—as extraterrestrial beings living in a virtual universe. P6 explained, \textit{``In their anniversary letter, the members admitted they had thought about quitting music and were hesitant about debuting as virtual idols. Choosing to wear `masks' and pursue their dreams really shows how serious they are.''}

Off-the-record anecdotes shared through the fan-artist chatting platform `Bubble' ~\cite{dearu_bubble} also intrigued participants with quirky details about their `backstage' lives, blending their fictional narratives with their ordinary human traits. P19 fondly recalled, \textit{“Bambi once showed us a hand-drawn sketch of his ‘plant friend,’ a pineapple cactus, proudly talking about how its arms had grown. It was so adorable.”} While seemingly mundane, these moments provided participants with a peek into the performers' everyday lives, grounding the extraordinary personas in relatable experiences.

As participants repeatedly encountered such seamful moments and PLAVE’s real identities became increasingly visible, they developed a curiosity about the individuals behind the avatars, despite PLAVE’s agency enforcing a strict prohibition against disclosing the performers' real identities. Unofficial personal details about the human performers became accessible as others had already identified them using subtle cues such as voice and other characteristics. Consequently, online information revealing these details became widespread.

As a result, the majority of participants (n=21, excluding P8, P9, and P16) were aware of details about the performers’ real identities at the time of the study, including their appearances and personal backgrounds. This awareness came through various channels: 11 participants encountered content featuring human performers through algorithmic recommendations on platforms such as YouTube, TikTok, X, and Google; 7 actively searched for this information themselves, while 3 learned about it through friends.

% As such, participants began to search for more information about the real individuals behind the avatars, although PLAVE's agency enforces a strict prohibition against disclosing the performers' real identities~\cite{vlast2024}. Leveraging subtle cues like voices, participants accessed online content revealing personal details about the human performers. This dissemination often stemmed from others piecing together and sharing such cues. P14 explained, \textit{``The algorithm frequently suggests content related to the human performers, making them more visible. Even while scrolling through Reels, it’s impossible not to notice since their voices are the same.''} 

% As a result, the majority of participants (n=21, except P8, P9, and P16) were aware of details of the performers’ real identities at the time of the study, including human performers' actual appearances and personal backgrounds. This awareness came through various channels. 11 participants reported encountering content featuring human performers through algorithmic recommendations on platforms such as YouTube, TikTok, X, and Google. 7 searched them by themselves, and 3 were informed by their friends. 

To summarize, participants’ interest in PLAVE evolved from merely recognizing the presence of human performers behind the avatars to gradually becoming aware of their real identities.

\subsection{Embracing or Detaching Seams} 
After acknowledging the human performers, participants showed distinct approaches to reconciling PLAVE's virtual and real identities in their fan engagements. Some participants integrated the real identity of the performers' real identities into their fan engagements, while others chose to disengage from it, focusing solely on the virtual personas. At the time of the study, more than half of the participants (n=14) embraced PLAVE's real identities as part of their engagement. In contrast, the remaining participants (n=10) deliberately disconnected from the real identities of PLAVE members, choosing to engage solely with their virtual representations.

\subsubsection{Embracing Seams: Bridging Virtual and Real Identities} %Integrating seams 

Fourteen participants reported actively incorporating PLAVE’s real identities into their engagement, seeing these seams as critical elements that bridged the gap between PLAVE’s virtual personas and the human performers behind them. Specifically, they felt these seams helped them \textbf{recognize the distinct personalities and charms of each member}. They saw seams provided a unique opportunity for them to explore and appreciate the nuances between PLAVE’s virtual and real identities, gradually discovering similarities and differences. When a member displayed a noticeable contrast between their virtual and real personas, they found more ways to appreciate the multifaceted nature of their identity. For instance, P3 described how she found amusement in understanding the contrast between a member’s dual identity:

\begin{quote}
\textit{It’s amusing because I can personally spot the differences, depending on how much I know about the real person. For example, Noah’s real self has a masculine, tough personality, but in the virtual world, everyone calls him a 'princess' because he has a soft side as well. As a result, he adapts to match the avatar and sincerely introduces himself as a princess. I find that so hilarious, but at the same time, it’s quite endearing and touching.} 
\end{quote}

Beyond recognizing individual traits, seams also enabled them to \textbf{contextualize PLAVE’s virtual activities} by integrating information from the performers' personal stories. They recalled instances where PLAVE members disclosed aspects of their daily lives. While these moments might have seemed incongruent with PLAVE’s virtual concept, they felt that such disclosures actually enhanced their overall understanding. P22 illustrated this by sharing how real-world cues helped her connect with PLAVE’s narratives:

\begin{quote}
\textit{When the members shared their travel stories in live streaming as PLAVE, I could recall the real-person performers' Instagram photos and stories. I then realized that those moments were 'the moments.'}
\end{quote}

Moreover, revealing aspects of daily life enhanced a sense of realism and authenticity, \textbf{reducing the psychological distance} between members and fans. PLAVE’s openness in sharing personal experiences fostered a deeper sense of closeness: ''\textit{It was adorable and made me feel closer to them}'' (P19). Similarly, participants noted that learning about the performers’ pre-debut history helped them better connect with and empathize with PLAVE’s journey. P4 emphasized how understanding both the real performer and the avatar was essential for fully engaging with PLAVE’s emotional depth:

\begin{quote}
\textit{To fully immerse yourself in PLAVE, you need to consider both the real-life performer and the avatar simultaneously. For example, when PLAVE recently cried while reading a letter to their past selves from a year ago, it wouldn’t be completely understandable if you only knew them as virtual idols. To fully grasp the moment, you need to understand the real performer’s journey leading up to this point.}
\end{quote}

For some participants who strongly embraced PLAVE’s real identities, seams facilitated a holistic integration of the group's dual personas, leading to fan activities that extended beyond the virtual realm. These participants engaged with the human performers as distinct individuals, separate from PLAVE’s official virtual identity. They reported watching the performers’ independent live streams, creating fan content focused on the performers themselves rather than their roles as PLAVE members, and following their personal Instagram and YouTube accounts. Offline activities also played a significant role, with fans attending the human performers’ concerts and fan meetings. P19 described how recognizing the connection between the real and virtual performers led her to engage more deeply with both:

\begin{quote}
\textit{After I realized the real performers' voices were identical in PLAVE's live streams, I gradually started listening to the human performers' songs, went to their fan meetings, and realized that their actions, words, and mindset were exactly the same as what PLAVE showed during the live streams. So, my initial resistance disappeared, and now I stan both equally.} 
\end{quote}

\subsubsection{Disengaging with Seams: Preserving the Virtual Fantasy}
However, not all participants actively embraced seams in their engagement with PLAVE. Ten participants chose not to engage with the real identities of the performers. While most of them (n=7) were already aware of who the real individuals were behind the avatars, they ceased to follow the personal updates of the human performer at the time of the study. A smaller proportion of participants (n=3) had not tried to search for the real identities of the performers, although they knew information about the real identities was easily findable online. 

These three participants \textbf{prioritized respecting the members’ virtual personas} over satisfying their curiosity about their real identities. P8 explained, \textit{``It's not that I don't want to know their real identity. I am curious about who they are, what they think, and what they like. But I want to respect their choice of how they present themselves, and it feels appropriate not to delve further into their real identities.''} For these participants, maintaining the distance from the human performer allowed them to continue engaging with the virtual personas without breaking the safety distance.

Meanwhile, seven participants expressed \textbf{discomfort or a sense of distance} after encountering the real performers behind the avatars. Particularly, these participants expressed confusion when they found the discrepancy between information from the human performer and the virtual personas. P6 experienced distraction from immersing herself in a seamless engagement with the virtual identity of PLAVE when she encountered a member presenting different information from her knowledge about the real person. She recalled, \textit{``I felt like he was acting, which created a sense of disconnect for me. It makes me wonder, what's real? How much of it is just an act?''}

These participants also experienced \textbf{disruption in engagement} when there was a notable gap between the virtual, unrealistic appearance of the avatar and the actual look of the real person that they learned from social media. For instance, P2 reported that her immersion into PLAVE was significantly disrupted: \textit{``When seeing that the actual performers’ height rankings differed from those of the avatars, I wondered, `why is their eye level different if they're actually the same height?'''} Also, the real appearance of the people behind the avatars disappointed some participants since it unveiled the fantastical male figures that the virtuality of PLAVE presented. 

\begin{quote} 
\textit{After becoming interested in PLAVE, I looked into their real identities because I wanted to know more about them. But when I found out, I felt suddenly disappointed. They were just like typical men in their 20s in real life. Once I realized that, I lost interest quickly.} (P6)
\end{quote}

This experience led some participants to actively block or avoid consuming information related to the human performers. P2 also explained, \textit{``The kind of appearance I prefer aligns more with the virtual side, and recently, I’ve become more interested in the virtual personas. Seeing the human performers has started to feel off-putting, almost like I’m witnessing a double life. So now, I’ve blocked all information about the human performers and actively avoid looking at anything related to them.''}

Similarly, participants who intentionally avoided investigating the actual people of PLAVE when encountering the seams (n=3, P8-9, 16) also expressed similar concerns. They worried that they might \textit{``become disenchanted’’} (P8) or feel disappointed if PLAVE’s real identities did not align with their expectations and imagination. For example, P16 mentioned: \textit{``I try not to look into their real identities because I want to enjoy stanning PLAVE without it feeling overwhelming. If I find out who they really are, it would completely break my immersion.''} (P16)

\begin{figure*}[!ht]
  \centering
  \includegraphics[width=480px]{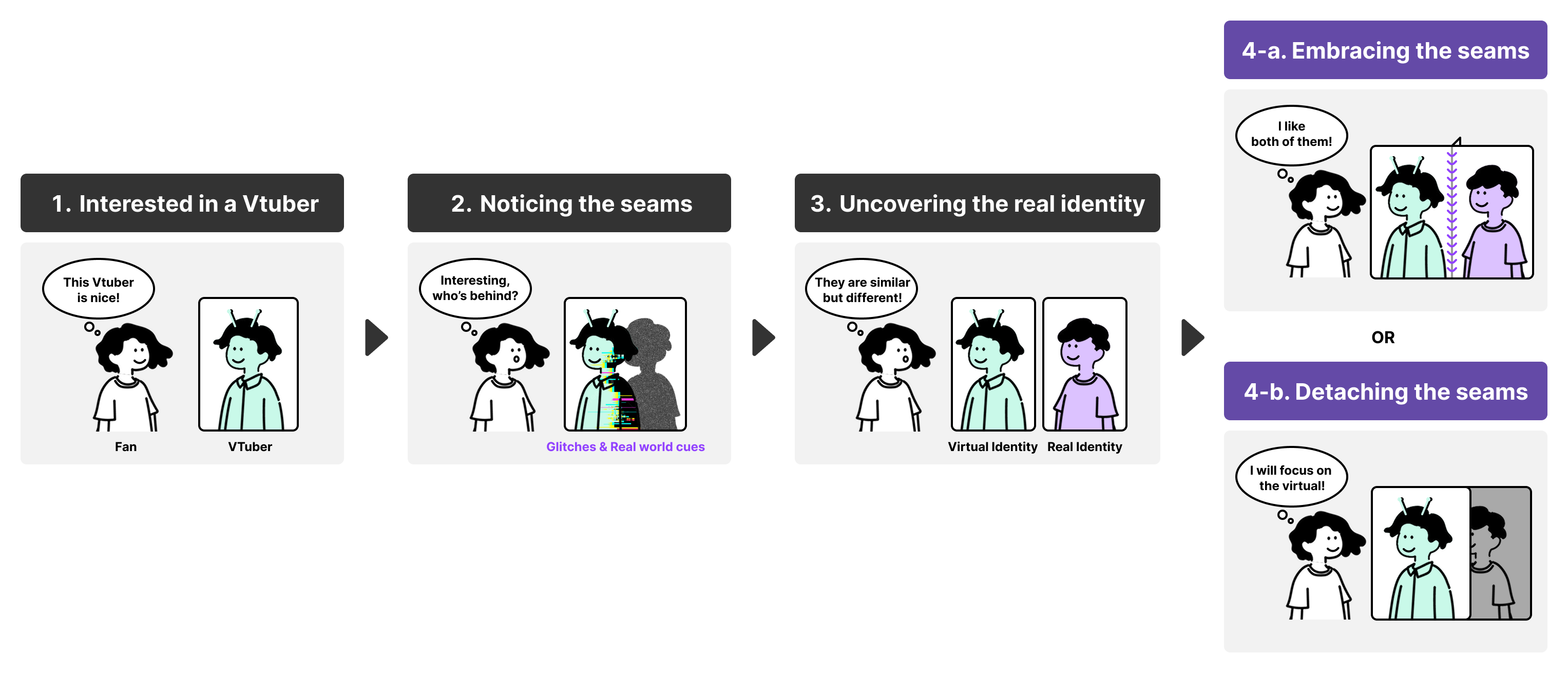}\captionsetup{justification=centering}
    \caption{Overview of our findings: A fan’s experience of engaging with a VTuber, progressing to recognizing their real identity, and ultimately choosing to embrace both personas or focus solely on the virtual.}
  \Description{This figure shows an overview of our findings focusing on fans' journeys around seams. A fan first becomes interested in a VTuber as a digital persona. As they engage further, they notice subtle cues hinting at the real person behind the avatar. This leads to the discovery of the real identity, where they recognize both similarities and differences between the virtual and real personas. In response, fans either embrace both identities as interconnected or detach from the real identity, choosing to focus solely on the virtual persona.}
  \label{fig:overview}
\end{figure*}

\section{Discussions}
Our study investigates how VTuber fans identify and perceive seams in VTubing and how these moments influence their overall engagement, using PLAVE as a case study.  By identifying technical glitches and implicit disclosures of their real identities as key seamful moments, our findings reveal the dynamic ways fans engage with and interpret the multi-layered identities of VTubers. Participants’ prior experiences with other VTubers and idols shaped their initial perceptions of PLAVE, while their evolving interactions with seamful moments sparked curiosity about the real individuals behind the avatars. These moments often acted as critical turning points, leading participants to either integrate virtual and real identities or disengage from real identities to maintain their focus on the virtual personas.

As such, our findings highlight the contrasting roles of seams in VTubing engagement. Seams could enhance fan engagement by fostering authenticity and deeper emotional connections, yet they could also disrupt immersion by exposing inconsistencies or misalignments. This duality underscores the need to understand seams not merely as disruptions but as integral elements of the VTuber-fan interaction, shaping how fans navigate and connect with the virtual and real-world dimensions of their favorite creators. These insights contribute to a better understanding of the broader VTubing ecosystem by emphasizing the importance of careful platform and content design that empowers users to navigate seams according to their preferences. 

In the following sections, we expand on the implications of our findings, focusing on the analytical potential of seams, their impact on user engagement, and proposing design strategies to address diverse fan needs around seams.

\subsection{Exploring Multi-layered Interactions with VTubers through a Lens of Seams}

Drawing from prior work that conceptualizes seams as moments of inconsistency or gaps within user experiences \cite{chalmers2003seamfulandseamless}, we initially adopted this view to describe the breakdowns between VTubers' virtual and real identities. However, our findings revealed that seams are not solely moments of disruption but also serve as a unique resource for fan engagement. This dual role aligns with alternative conceptualizations of seams as moments where breakdowns both disrupt and illuminate underlying complexities and user-system dynamics \cite{sengers2006staying,erickson2016infrastructuring,rubambiza2022seamless}. 
 % These studies highlight the duality of seams as disruptions that also foster reflection, adaptation, and meaningful engagement \cite{vertesi2014seamful,chalmers04seamfulinterweaving}, an interpretation that closely aligns with our findings on the role of seams in VTuber fan interactions.

This nuanced role of seams has been largely underexplored in the context of VTuber-viewer interactions. Prior research has often focused on VTubers’ self-presentation and the creation of detached personas, emphasizing their ability to congruent virtual identities \cite{suan2021performing,schmieder2024waiting,bredikhina2022becoming}. In contrast, our findings highlight seams as pivotal moments that simultaneously disrupt and/or deepen engagement, shaping how fans interpret and connect with the multi-layered identities of VTubers.

These insights suggest the need to reconsider seamlessness as the default objective in VTubing system design \cite{inman2019beautiful}. By intentionally embracing and designing for seamful interactions, platforms may be able to enrich fan engagement through greater transparency, authenticity, and opportunities for personalization. Allowing fans to navigate these moments based on their individual preferences could foster more diverse and adaptive user experiences.

In conclusion, our study demonstrates the value of seams as a conceptual lens for understanding VTuber-fan engagement. Seams function as both disruptions and meaningful moments, facilitating a dynamic interplay between virtual and real identities and empowering fans to navigate these complexities in alignment with their personal preference.

\subsection{Roles of Seams in Constructing Fan Experiences}

In this section, we expand on these findings to explore the broader roles that seams could play in the VTubing ecosystem, considering their potential to enhance or challenge fan experiences and their implications for platform and content design.

\subsubsection{Seams as Catalysts for Deeper Engagement and Authenticity}

Based on our findings, seams serve as powerful invitations for viewers to engage more deeply by revealing the individuality of the human performers behind the avatars. Participants encountered these seamful moments through technical glitches or implicit disclosures during live streams, which exposed the performers’ distinct personalities and added layers of familiarity and charm. Similarly, personal disclosures—such as anecdotes about daily life or reflections on pre-debut experiences—helped contextualize the crafted virtual narratives, blending virtual and real interactions into a cohesive and engaging understanding of VTubers’ multi-layered identities.

This finding builds on existing research on the strategic use of seams to capture viewer interest \cite{wijaya2023language,wan2023investigating}. However, our study demonstrates that these moments do more than simply intrigue viewers—they transform passive consumption into active, empathetic engagement. By providing a window into the human elements behind virtual personas, seams allow viewers to see beyond avatars’ polished exteriors, fostering intimacy. This suggests that seams may serve as a critical mechanism for audience engagement, inviting them to navigate the interplay between the virtual and real dimensions of VTubers. These interactions highlight seams’ potential to deepen fan loyalty and investment, offering a richer, more multifaceted VTubing experience.

Beyond enhancing engagement, our findings suggest that seams also shape the perception of authenticity in VTubing. Prior research emphasizes that authenticity is primarily tied to the coherence of virtual personas, rather than the real-world individuals behind them \cite{schmieder2024waiting, lu2021more, aicher2023influence, huang2024avatar}. However, our findings extend this view by demonstrating that authenticity in VTubing can also emerge from integrating real-world elements, such as personal anecdotes, pre-debut histories, and daily life experiences. This suggests that seams could serve as a design mechanism for fostering authenticity in virtual personas, offering a way to balance immersion with personal connection. Furthermore, this highlights the evolving nature of authenticity, where the interplay between the virtual and real continually reshapes user expectations and engagement.

\subsubsection{Seams as Disruption}
Despite their positive aspects, seams also posed challenges to fans' engagement with VTubers. While personal disclosures were often appreciated, subtle discrepancies between virtual and real-world personas, such as unexpected glimpses of physical differences or inconsistencies in personal details, disrupted some participants' immersion. These unanticipated moments shifted fans' focus away from the virtual personas, causing temporary detachment. Some participants even expressed concerns about disillusionment upon recognizing the real identities of PLAVE members. This highlights the delicate nature of seamfulness in VTuber-viewer interactions, revealing the intricate dynamics of performing dual identities.

As such, unpredictable encounters with seams can significantly compromise audience engagement. Drawing on Kratz and Ballagas' \cite{kratz2009unravelling} categorization, the impact of seams can be understood through their predictability. ``Random seams,'' characterized by inconsistent and unpredictable manifestations, challenge users’ ability to adjust their behaviors and expectations. In contrast, ``regular seams,'' which appear in consistent patterns, are easier to anticipate, allowing users to make sense of and adapt to them.

Given this, strategically modulating seams would become critical in VTubing. By presenting identity disclosures in a consistent and predictable manner, content creators may reduce disruptions caused by unexpected discrepancies and enhance the reliability of their personas. Considering that the disclosure of real identities is often inevitable in the VTubing experience, managing seams thoughtfully would mitigate negative impacts and sustain audience engagement.

\subsection{Toward the Artful Management of Seams in VTubing Experiences}

Our findings suggest that seams in VTubing are neither inherently good nor bad; instead, they exist on a continuum that users navigate based on their engagement preferences, as proposed by Kratz and colleagues \cite{kratz2009unravelling}. Fans strategically balance seamful and seamless interactions to align with their specific engagement goals. This suggests that platform design should focus on enabling this \textbf{artful management} rather than attempting to completely erase or amplify seams \cite{vertesi2014seamful, rubambiza2022seamless}. Ultimately, VTubing platforms should be designed with greater adaptability, ensuring that their features and functionalities are appropriately contextualized and responsive to the diverse ways users engage with VTuber content.

Despite the need for artful management of seams, we also observed instances where these differing preferences were overridden by uncontrollable factors embedded in platforms, such as algorithmic recommendation systems. For example, YouTube's recommendation systems inadvertently exposed videos featuring real-person performers to participants’ newsfeeds, even though they did not wish to learn about their real identities. Sometimes, this exposure was made maliciously to contempt VTubers, making their intimate engagement with PLAVE vulnerable. Unfortunately, users have little control over avoiding sudden exposure to real personas, except through self-developed tactics \cite{park2024exploring}, such as individually blocking videos featuring real performers.

To address this challenge, platforms would need to empower users with the ability to manage their engagement with seams. For example, unexpected exposure to sensitive information regarding VTubers’ real identities can be bypassed by implementing personalized settings that accommodate individual users' preferences and expectations regarding seam exposures. For instance, platforms could introduce a more granular tagging system for VTuber-related content ~\cite{peng2010collaborative,park2024exploring}. This system may include categories such as `Hamin's real identity,' `behind-the-scenes,' and `virtual-centric,' thereby clearly delineating the nature of the video. Leveraging these tags, content recommendation algorithms could be calibrated to reflect users' seam exposure preferences. Specifically, based on user-defined settings, the algorithm could also modulate the frequency of recommendations for particular content types, ensuring that users experience seam exposure at a level that aligns with their comfort and engagement preferences. Additionally, platform guidelines that support comfortable VTubing experiences—free from unwanted or intrusive content—could further enhance viewer autonomy.

% Our study also highlighted the significant potential of seams in studying VTubers' viewer experiences, offering a liminal space to explore how viewers dynamically interact with the virtual and real worlds, as well as the intricate blending of VTubers' dual identities. By thoroughly analyzing the case of PLAVE as one of the instances of ``beautiful seams'' ~\cite{inman2019beautiful,lyytinen2002ubiquitous}, we identified unique design opportunities for viewers to artfully navigate seams and enhance their immersive VTubing experiences.

\subsection{Limitations and Future Work}
While our study offers valuable insights into the role of seams in shaping viewers' interactions with VTubers, it is important to acknowledge its limitations and suggest directions for future work.

First, the majority of our interview participants were women in their 20s and 30s, reflecting the dominant demographic of PLAVE’s fan base. While this focus aligns with the group’s primary audience, it may limit the generalizability of our findings to other demographic groups. Additionally, our study did not capture perspectives from individuals who initially liked PLAVE but later disengaged or those who were unable to immerse themselves in PLAVE’s content. As such, future studies should aim to broaden the participant pool across more diverse genders, age groups, cultural contexts, and levels of fan engagement. Expanding the scope to include these perspectives could offer a more holistic view of how seams supported VTuber-viewer interactions.

Second, our findings are situated within the context of PLAVE, a virtual idol group operating with highly sophisticated VTubing technology, including advanced motion-capture algorithms. These technological advancements, combined with their framing as idols within the broader K-pop culture, likely enhanced fan immersion and engagement. As a result, our findings may not fully apply to VTubers operating in less technologically advanced settings or those framed outside of the K-pop idol narrative. Future research should explore how the role and perception of seams differ across varying technological capabilities and narrative frameworks, offering insights into the generalizability of these findings.

Furthermore, future studies could incorporate the perspectives of performers in managing seams while also highlighting the invisible labor of human performers behind the scenes \cite{cheon2024creative, harvey2024cadaver}. Addressing these additional layers could provide deeper insights into the intricate dynamics of identity negotiation within VTubing ecosystem.

\section{Conclusion}
This study examined how VTuber fans recognize and interpret seams—moments where the boundary between a VTuber’s virtual and real identities blurs—and how these moments influence their engagement with VTubers, using PLAVE as a case study. Our findings revealed that participants gradually became aware of seams through technical glitches and implicit disclosures of real-world identities during live streams. Participants exhibited distinctive responses to PLAVE’s virtual and real identities, with some actively embracing the performers' real-world presence, while others prioritized their virtual personas to maintain immersion. Seams, at times, disrupted engagement by making discrepancies between members' dual identities more apparent, leading to moments of detachment. However, they also enhanced fan experiences by offering opportunities to appreciate the relatable and authentic aspects of the human performers, fostering a sense of connection and deeper engagement. Based on these findings, we underscore the dual role of seams in VTubing—both as potential disruptors of immersion and as mechanisms for deeper engagement.
Overall, our findings highlight the importance of platform design that enables users to manage seam exposure, fostering more adaptive and personalized interactions with VTuber content.

\begin{acks}
We sincerely appreciate our participants and reviewers for their significant contributions to this work. This research is supported by the SNU-Global Excellence Research Center establishment project and the Undergraduate Researcher Program (URL) in Information Science and Cultural Studies at Seoul National University.
\end{acks}

\balance

\bibliographystyle{ACM-Reference-Format}
\bibliography{ref}
\end{document}